\documentclass[12pt]{iopart}
\usepackage{graphicx}
\usepackage{amssymb}
\usepackage{color}
\usepackage{cite}
\begin{document}

\title{A possibility of high spin hole states \break in doped CoO$_2$ layered systems}

\author{      Krzysztof Ro\'sciszewski$^1$ and
              Andrzej M. Ole\'s$^{1,2}$  }

\address{$^1$ Marian Smoluchowski Institute of Physics, Jagellonian
              University, \\ Reymonta 4, PL-30059 Krak\'ow, Poland }
\address{$^2$ Max-Planck-Institut f\"ur Festk\"orperforschung, \\
              Heisenbergstrasse 1, D-70569 Stuttgart, Germany }
\ead{krzysztof.rosciszewski@uj.edu.pl; a.m.oles@fkf.mpg.de}

\date{\today}

\begin{abstract}
We introduce and investigate an effective five-band model for $t_{2g}$ 
and $e_g$ electrons to describe doped cobalt oxides with Co$^{3+}$ and 
Co$^{4+}$ ions in two-dimensional CoO$_2$ triangular 
lattice layers, as in Na$_{1-x}$CoO$_2$.
The effective Hamiltonian includes anisotropic kinetic energy (due to 
both direct Co-Co and indirect Co-O-Co hoppings), on-site Coulomb
interactions parameterized by intraorbital Hubbard repulsion $U$ and
full Hund's exchange tensor, crystal-field terms and Jahn-Teller static
distortions. We study it using correlated wave functions on $6\times 6$
clusters with periodic boundary conditions. 
The computations indicate low $S=0$ spin to high $S=2$ spin abrupt 
transition in the undoped systems when increasing strength of the 
crystal field, while intermediate $S=1$ spins are not found. Surprisingly,
for the investigated realistic Hamiltonian parameters describing low 
spin states in CoO$_2$ planes, doping generates high $S=\frac{5}{2}$ 
spins at Co$^{4+}$ ions that are pairwise bound into singlets, seen 
here as pairs of up and down spins. It is found that such singlet pairs 
self-organize at higher doping into lines of spins with coexisting 
antiferromagnetic and ferromagnetic bonds, forming stripe-like 
structures. The ground states are insulating within the investigated 
range of doping because computed HOMO-LUMO gaps are never small enough.
\\
{\it Published in: J. Phys.: Condensed Matter} {\bf 25}, 345601 (2013) [IOP Select].
\end{abstract}


\submitto{\JPCM}

\maketitle

\section{Introduction}

Cobalt oxides are quite unique due to large splitting between $t_{2g}$
and $e_g$ states in their electronic structure and a nonmagnetic ground
state distinguishing LaCoO$_3$ from a conventional Mott insulator
\cite{Ima98}. In the singlet state ($S=0$) all $t_{2g}$ states are
filled at each Co$^{3+}$ ion, and any magnetic or orbital order is
excluded. With increasing temperature this compound undergoes a
spin-state transition from a nonmagnetic ($S=0$) to intermediate spin
($S=1$) state \cite{Kor96}. In the latter $t_{2g}^5e_g^1$ state the 
orbital $e_g$ degree of freedom is released and the ordered $e_g$ 
orbitals support $A$-type antiferromagnetic (AF) order, similar to that 
observed in LaMnO$_3$ \cite{Dag01,Kov10}. This type of order follows in 
LaMnO$_3$ from the spin-orbital model designed for high spin $(S=2)$ 
states of Mn$^{3+}$ ions in $t_{2g}^3e_g^1$ configuration \cite{Fei99}.

Transitions from low spin to high spin states in other cobalt oxides,
including two-dimensional (2D) CoO$_2$ triangular lattice layers such 
as in Na$_{1-x}$CoO$_2$, (Bi,Pb)-Sr-Co-O or Ca$_3$Co$_4$O$_9$ compounds,
have not been reported so far. The properties of Na$_{1-x}$CoO$_2$
systems change under doping, with an interesting interplay between the
magnetic order and superconductivity \cite{Foo04}. On the one hand, the 
electronic structure of these systems is of great interest and gives 
many interesting physical features that follow from intrinsic 
frustration of magnetic interactions on the triangular lattice. On the 
other hand, the cobalt ions Co$^{3+}$ in the undoped compounds, as for 
instance in NaCoO$_2$, are nonmagnetic \cite{deV05} in $t_{2g}^6$ 
configuration. This is in contrast to the majority of other cobaltates, 
with a more conventional three-dimensional (3D) structure, where 
intermediate or high spin states are found at cobalt ions (of various 
valence). However, also here doping leads to a radically different 
behaviour from that of a free charge embedded in a band insulator 
\cite{Kha08}.

Cobaltates with 2D triangular lattice were the subject if intense
research
\cite{Foo04,deV05,Kha08,bourg09,mizok04,bourg07,kroll06,koshi03,inder05,yamak10}
which led to the common view that nonmagnetic (undoped) parent
compounds develop on doping {\it low spin} symmetric superstructures, 
some of them ferromagnetic (FM), some of them AF, and some others 
stripe-like, as described in an excellent paper by Mizokawa \cite{mizok04}.
Because in the investigated substances the Co$^{3+}$ ions belonging to
undoped 2D triangular lattice layers are nonmagnetic which suggests that
the $e_g$ levels are unoccupied in $t_{2g}^6$ states and stay also
unoccupied when doping occurs. In the present paper we are verifying
this common view and we provide arguments that the $e_g$ levels play a 
prominent role there and could lead to high spin states of Co$^{4+}$ 
ions in some doped systems with triangular lattice. In this respect this
paper can be considered as a supplementary to an earlier study of the
eleven-band $d$-$p$ model in Na$_{1-x}$CoO$_2$ \cite{yamak10}.

The paper is organized as follows. In section 2 we introduce an 
effective model for 3d electrons and provide available information 
concerning its parameters. The model is next solved on $6\times 6$ 
clusters for several doping levels by self-consistent calculations 
based on the Hartree-Fock (HF) approach with electron correlations 
implemented by an exponential local ansatz, as explained in section 3. 
The numerical results are presented and analyzed in section 4.
The paper is concluded in section 5, where we also point out certain
possible experimental implications of the present studies.
Appendix presents the kinetic energy elements for Co-Co hopping in the 
effective model which includes only $3d$ orbitals at Co ions.

\section{The effective model for $3d$ electrons}

We investigate strongly correlated electrons in doped 2D monolayer with
triangular lattice occupied by Co$^{3+}$ ions or Co$^{4+}$ ions, such
as in Na$_{1-x}$CoO$_2$ or in (Bi,Pb)-Sr-Co-O compounds.
The effective model introduced below takes into account only effective 
$d$-type Wannier orbitals at Co sites (resulting from hybridization of 
$3d$ cobalt orbitals with surrounding oxygen $2p$ orbitals). The
hybridization is responsible for renormalisation of the bare cobalt
Hamiltonian parameters used below in the effective Hamiltonian.
Note that formally one can obtain such an effective Hamiltonian by a 
procedure of mapping Hartree-Fock or local density approximation with 
Coulomb interaction $U$ (LDA+$U$) results, obtained in a multiband model 
featuring cobalt and oxygen orbitals. Such a mapping should preserve the 
structure of lowest energy levels.

When the splitting between $t_{2g}$ and $e_g$ states is large, it might
be argued that a three-band model including $t_{2g}$ orbitals only 
could be sufficient to describe the Co$^{3+}$ ions in the {\it low spin} 
state ($S=0$) in Na$_{1-x}$CoO$_2$ or (Bi,Pb)-Sr-Co-O compounds
\cite{mizok04,kroll06,bourg07,bourg09}. Here we study a five-band
model to obtain a better insight into:
($i$) the effects of doping, i.e., the consequences of introducing
Co$^{4+}$ ions into the parent (undoped) system with Co$^{3+}$ ions;
($ii$) the crossover regime when due to hypothetical crystal field 
weakening (and therefore smaller distance between $t_{2g}$ and $e_g$ 
levels) one may expect a transition from low spin to high spin Co ions
in the ground state.

The effective Hamiltonian for a 2D triangular lattice of Co ions 
consists of four parts:
\begin{equation}
{\cal H}= H_{\rm kin}+H_{\rm cr1}+H_{\rm cr2}+H_{\rm JT}+H_{\rm intra},
\label{model}
\end{equation}
The kinetic (hopping) part of the Hamiltonian is:
\begin{equation}
H_{\rm kin}=   \sum_{ \{i j \mu \nu\} \sigma} t_{ i \mu, j \nu}
 d^{\dagger}_{i \mu \sigma}  d_{j \nu \sigma}\,,
\label{Hkin} 
\end{equation}
where $d_{j\nu\sigma}$ denotes electron annihilation operator at site 
$j$, $\nu = xy, yz, zx, x^2-y^2, 3z^2-r^2$ labels $3d$ orbitals, and 
$\sigma=\uparrow,\downarrow$ corresponds to up and down electron spin. 
The nonzero hopping elements describe both indirect
cobalt-oxygen-cobalt transitions, and direct cobalt-cobalt hoppings.
They are given by two parameters, $t_0$ and $t_1$, and defined in 
table 1 and table 2 presented in Appendix. 

The lattice is characterized by the lattice vectors 
(we take a lattice constant $a=1$)
\begin{equation}
\label{vectors}
\mathbf{a}_1=\frac{1}{\sqrt{2}}(0, 1,-1)\,, \hskip .5cm
\mathbf{a}_2=\frac{1}{\sqrt{2}}(-1,0, 1)\,, \hskip .5cm
\mathbf{a}_3=\frac{1}{\sqrt{2}}(1,-1, 0)\,.
\end{equation}
These vectors are presented in figure 1 by Indergand {\it et al} 
\cite{inder05}. The parameters of indirect (effective) hoppings 
$\propto t_0$ can be obtained from the analysis of a multiband $d$-$p$ 
model --- they obey Slater-Koster rules \cite{livermore} and follow from 
the lowest order perturbation theory \cite{Zaa93}. For instance, using 
tight binding formalism we obtain for a $t_{2g}$ system \cite{koshi03} 
that the hopping amplitude from $xy$ orbital at site number 0 via $p_x$ 
oxygen orbital to $zx$ orbital at nearest neighbor site along the 
lattice vector $\mathbf{a}_1$ is equal $t_0 = P^2_{pd\pi}/\Delta$; here 
we use this hopping as a unit and take $t_0=0.3$ eV. More details on 
the possible choice of the microscopic parameters which justify this 
value and o n finite hopping elements are given in Appendix, see table 1.

The direct cobalt-cobalt hoppings are parametrized by the element 
$t_1\equiv\frac{1}{2}P_{dd\pi}$, being the hopping between two 
neighbouring $t_{2g}$ orbitals lying in the plane perpendicular to the 
bond direction ${\bf a}_n$ (for instance two $xy$ orbitals for a bond 
along ${\bf a}_1$); here $t_1=0.05$ eV. All $d$-$d$ hopping elements
are collected in table 2 in Appendix.

Simplified Jahn-Teller (JT) part of the Hamiltonian was proposed by
Toyozawa and Inoue \cite{toyoz66} for $e_g$ and for $t_{2g}$ deformations:
\begin{eqnarray}
\label{HJT}
H_{\rm JT}&=&
 \frac{1}{2}\,\sum_i \Big\{K_{\rm br} Q_{1i}^2  +K_{\rm JT}
 \left[ Q_{2i}^2+ Q_{3i}^2 + Q_{4i}^2+ Q_{5i}^2  + Q_{6i}^2\right]\Big\}\nonumber\\
&+&g_{\rm JT}  \sum_i \Big\{
-Q_{1i}( n_{i, x^2-y^2} + n_{i, 3z^2-r^2})  \nonumber \\
&+&Q_{2i} \sum_\sigma ( d^{\dagger}_{i, x^2-y^2 \sigma}  d_{i,3z^2-r^2,\sigma} +
d^{\dagger}_{i, 3z^2-r^2, \sigma}  d_{i, x^2-y^2,\sigma} ) + \nonumber\\
&+& Q_{3i} \sum_\sigma
( d^{\dagger}_{i, x^2-y^2,\sigma}  d_{i, x^2-y^2,\sigma} -
d^{\dagger}_{i, 3z^2-r^2 ,\sigma}  d_{i, 3z^2-r^2,\sigma} ) +  \nonumber\\
&+& Q_{4i} \big( d_{i, xy,\sigma}^\dagger d_{i, zx,\sigma}
+d_{i, zx,\sigma}^\dagger d_{i, xy,\sigma} \big)  \nonumber \\
 &+&  Q_{5i} \big( d_{i, xy,\sigma}^\dagger d_{i, yz,\sigma}
 +d_{i, yz,\sigma}^\dagger d_{i, xy,\sigma} \big) \nonumber \\
&+& Q_{6i} \big( d_{i, yz,\sigma}^\dagger d_{i, zx,\sigma}
+d_{i, zx,\sigma}^\dagger d_{i, yz,\sigma} \big)\Big\}
\end{eqnarray}
where $ Q_{1i},..., Q_{6i}$ denote static JT deformations of the $i$-th
CoO$_6$ octahedron. To make the model \eref{model} as simple as possible 
we assume that:
($i$) the same set of parameters $\{K_{\rm JT},g_{\rm JT}\}$ is suitable 
for $e_g$ ($Q_{2i},Q_{3i}$) and for $t_{2g}$ ($Q_{4i},Q_{5i},Q_{6i}$)
modes, and 
($ii$) the breathing mode $Q_1$ can be neglected  $K_{br}/K_{JT}\gg 1$
(note that in manganites $K_{\rm br}/K_{\rm JT}\approx 2$ \cite{Dag01}). 
These simplifying assumptions allow one to make only qualitative 
predictions (concerning the JT effect), but any quantitative analysis 
would require more precise information about the coupling constants.

In manganites typical values for $g_{\rm JT}$ and $K_{\rm JT}$ are:
$g_{\rm JT}=3.8$ eV \AA$^{-1}$ and $K_{\rm JT}=13$ eV \AA$^{-2}$,
respectively (see \cite{bilayer} and references therein). For cobalt
oxides the values of $g_{\rm JT}$ and $K_{\rm JT}$ are not known. Here
we will arbitrarily assume the same
value of $K_{\rm JT}=13$ eV \AA$^{-2}$ and we estimate the value of
$g_{\rm JT}\approx 1.6$ eV \AA$^{-1}$ from some experimental data
reported in a different cobalt compound. Namely Pradheesh {\it et al}
\cite{gJT} reported strong $Q_3$ JT distortion in CoO$_6$ octahedra,
i.e., two long Co-O (apical) bonds and four shorter Co-O bonds when the
central cobalt was Co$^{3+}$ ion with an intermediate spin. From their
data ($Q_3\approx 0.12$ \AA $\;\;$) we make a  jump to our
(different) systems and the crude estimate follows
$g_{\rm JT}\approx 1.6$ eV\AA$^{-1}$, i.e., the value by half smaller
than the one in manganites.

Crystal field part of the Hamiltonian consists of two parts,
$H_{\rm cr1}$ and $H_{\rm cr1}$. The first one is responsible for the
splitting within the group of $t_{2g}$ levels,
\begin{equation}
H_{\rm cr1}  \propto\frac{1}{3}\sum_i
\left(d_{i,xy,\sigma}^\dagger+ d_{i,yz,\sigma}^\dagger+d_{i,zx,\sigma}\right)
\left( d_{i,xy,\sigma}^{}+ d_{i,yz,\sigma}^{}+d_{i,zx,\sigma}^{}\right).
\end{equation}
Namely, the orbital
$\big(|xy\rangle+|yz\rangle+|zx\rangle\big)/\sqrt{3}$ is placed below
two degenerate states: 
$\left(|xy\rangle+e^{\pm\frac{2\pi i}{3}}|yz\rangle
+e^{\pm\frac{4\pi i}{3}}|zx\rangle\right)/\sqrt{3}$. According to 
Bourgeois {\it et al} \cite{bourg07}, $H_{\rm cr1}$ can be reexpressed 
in the form
\begin{equation}
H_{\rm cr1} = -D_1 \sum_i\sum_{\alpha\neq\beta}^{'}
d_{i\alpha,\sigma}^\dagger d_{i\beta,\sigma}^{},
\end{equation}
where the summation $\sum_{\alpha \neq \beta}^{'}$ runs only over
$\{xy, yz, zx\}$ orbitals. The magnitude of splitting amounts to
$3 D_1\simeq 0.315$ eV, following the results of {\it ab initio}
cluster computations performed for Na$_{1-x}$CoO$_2$ compounds and 
other available estimates \cite{bourg07,kroll,pilla08}.
We believe that this part of Hamiltonian does not influence the results
of the present investigation in any significant way --- anyway we 
include it to be consistent with other models used in this field.

The second (simplified) part of crystal field Hamiltonian modeling 
depends on the splitting between $t_{2g}$ and $e_g$ orbitals and can 
be expressed using site occupations operators
$n_{i\alpha}=\sum_\sigma d_{i,\alpha,\sigma}^\dagger d_{i,\alpha,\sigma}$
as follows
\begin{equation}
H_{\rm cr2} =  D_2\sum_i\left( n_{i,x^2-y^2} +  n_{i,3z^2-r^2}
-n_{i,xy}  -n_{i,yz} - n_{i,zx} \right),
\end{equation}
where the (experimental) magnitude of the splitting is large for low
spin cobaltates. The following values were suggested: $2.5$ eV
\cite{koshi03}, $1.7$ eV \cite{zou04} and $1.5$ eV \cite{zaliz00}.
In addition, the experimental splitting is strongly dependent on doping
\cite{kroll06,huang04}. For other kind
of cobalt oxides (i.e., not the ones studied in the present paper) Merz
{\it et al} \cite{merz11} claim that crossover between low spin cobalt
($S=0$) and intermediate/high spin cobalt oxides occurs at
$\approx 1.0-1.4$ eV. How these estimates are related to the model
Hamiltonian value of $D_2$ is difficult to say. The naive estimation
(when taking into account only $H_{\rm cr2}$ and $H_{\rm intra}$)
is that on-site experimental splitting (between single ion configuration 
with six $t_{2g}$ electrons and paramagnetic $t_{2g}^5e_g^1$ 
configuration) is equal $2D_2-5B$ where $B$ is Racah parameter (see 
$H_{\rm intra}$ below and the following comments).
This naive estimate does not take into account correlations and what
is more important does not take into account any kinetic effects
which arise when including $H_{\rm kin}$ and $H_{\rm cr1}$ into
consideration. Therefore, in this paper, at first $D_2$ will be
treated as a variable parameter but finally we will fix below 
a representative value of $D_2=1.25$ eV 
(if we accept $B$ = 0.1 eV then the level splitting is: $2D_2-5B=2$ eV).
Let us stress once again that the energy splitting between $t_{2g}$
and $e_g$ levels is a very important parameter. It is large for low 
spin compounds and small for the ones with intermediate/high spin 
states \cite{merz11,hollm08}.

Finally, the third type of splitting which occurs within $e_g$ levels
in the present model is neglected. The paramagnetic ground state in the 
undoped compound can show some magnetic features upon subsequent doping 
(due to rising concentration of Co$^{4+}$ ions). Let us quote a remark 
from the literature that for $g_{\rm JT}=D_1=t_1=0$ and $D_2\gg 1$ the 
description of electronic states can be based on a three-band model 
(with $t_{2g}$ orbitals $\{xy,yz,zx\}$ orbitals only) 
and four kagome sublattices \cite{koshi03,inder05}. 
For $0\neq D_1\gg t_0$ we have dispersionless single band model. 
In reality, however, both $t_0$ and $t_1$ play an important role and 
decide about the electron distribution and total spin in the ground state.

The last part of the Hamiltonian $H_{\rm int}$ is strong local on-site 
electron-electron interaction. Here we adopt a more general form of the
degenerate Hubbard model \cite{Ole84,Horsch}
\begin{eqnarray}
H_{\rm int} &=&
  U \sum_{i, \mu}  n_{i\mu,\uparrow} n_{i\mu,\downarrow}
+\frac{1}{2}\sum_{i,\mu\neq\nu}
\left(U-\frac{5}{2}J_{\mu\nu}\right)n_{i\mu}n_{i\nu}\nonumber  \\
&-& \frac{1}{4} \sum_{i,\mu\neq\nu}
J_{\mu\nu}\big( n_{i\mu\uparrow}-n_{i\mu\downarrow}\big)
    \big( n_{i\nu\uparrow}-n_{i\nu\downarrow} \big),
\label{hubbard-intra}
\end{eqnarray}
where $ J_{\mu\nu}$ is the tensor of on-site interorbital exchange
elements for $3d$ orbitals which can be expressed using Racah
parameters $B$ and $C$ \cite{Horsch,Griffith} (see table 1 given by
Horsch \cite{Horsch}). Note that each pair of different orbitals 
$\mu\neq\nu$ is included twice in equation \eref{hubbard-intra}. In 
simple situations when the system can be
described solely in terms of $t_{2g}$ orbitals all $J_{\mu\nu}=3B+C$
and define the unique Hund's coupling $J_H$. Another simple situation
is encountered when only $e_{g}$ orbitals are partly filled --- then 
all Hund'd exchange elements are again the same, and 
$J_H\equiv J_{\mu\nu}=4B+C$. Cross terms between
$t_{2g}$ orbitals and $e_{g}$ orbital are different and smaller (we
remind that for $J_{\mu\nu}$ we take the entries from table 1 in
\cite{Horsch}). Furthermore, in the present investigation for the sake
of simplicity we use the following empirical ansatz: $C=4B$ --- in the
literature it is frequently used (and is quite realistic \cite{Fei99}) 
for transition metal oxides with ions in various configurations $3d^n$.
However, this relation is only approximately satisfied in real 
compounds and some correlations might be necessary \cite{buene05}.

Some comments are necessary about simplifications we made in equation
\eref{hubbard-intra}. Namely the last term in equation \eref{hubbard-intra} 
results from mean field approximation done to original \cite{Horsch} 
spin-spin SU(2) scalar product: the spin symmetry is explicitly broken
and the quantization axis is fixed in spin space. Then, the full Hund's
exchange interaction is replaced by the Ising term. We neglect here the
spin-flip terms in Hund's exchange that go beyond the mean field 
approximation and could lead to spin-orbital entanglement which could be 
studied only in more sophisticated many-body
treatments \cite{Ole12}. However, this
approximation is commonly used, for instance in the LDA+$U$ approach,
because in the HF approximation (when applied to exact intraatomic
interaction) one obtains the same final result for electronic 
interactions. The second approximation we made here is the neglect of 
double occupancy transfers occurring due to Coulomb interactions between
two different orbitals \cite{Ole84}.
The penalty due to these approximations, we have to accept, is twofold:
($i$) neglect of some contributions when correlations
are included, using the HF states, and
($ii$) this approach excludes explicitly spiral-like spin arrangements;
they can not be properly described when using the approximate form of
equation \eref{hubbard-intra}. These approximations influence 
quantitatively but not qualitatively the multiplet structure \cite{Hor11}.
For a triangular lattice this turns out not to be a serious problem.

The on-site interaction Hamiltonian \eref{hubbard-intra} can be rewritten
using the electron density operators as follows:
\begin{eqnarray}
H_{\rm int} &=&
  U \sum_{i, \mu}  n_{i\mu, \uparrow} n_{i\mu, \downarrow}
 +\frac{1}{2}\sum_{i,\mu\neq\nu,\sigma} \big( U-3J_{\mu\nu}\big)
 n_{i\mu,\sigma} n_{i\nu,\sigma}  \nonumber  \\
 &+& \frac{1}{2} \sum_{i,\mu\neq\nu,\sigma} \big(U-2J_{\mu\nu}\big)
 n_{i\mu, \sigma}n_{i\nu,-\sigma}.
\label{hubbard2-intra}
\end{eqnarray}
The estimations of an {\it average} Hund's exchange $J_H$ are: 0.84 eV 
(close to atomic value) \cite{mizok95,zhang04,wakis08}; 0.72 eV 
\cite{kroll06} and even a value smaller than 0.7 eV \cite{zou04}. 
An effective value of $J_H=0.35$ eV deduced from an exact solution of a 
single CoO$_6$ cluster which includes strong $d$-$p$ hybridization and 
from fitting to x-ray absorption experiments \cite{bourg09}
is much smaller. A small value of $J_H=$ 0.28 eV follows also from
{\it ab initio} computations when exchange interaction is strongly
reduced (from the atomic value) by some screening effects \cite{land06}.
Such values have to be considered as semiempirical parameters while
larger values are appropriate when the correlation effects are treated
explicitly, as in the present paper. Here we adopt the value following
Kroll, Aligia and Sawatzky \cite{bourg09}, i.e.,
$J_H=B+4C\simeq 7B\simeq 0.7$ eV given by $B=0.1$ eV.

For Hubbard repulsion $U$ one finds 5.5-6.5 eV used in multiband HF
models by several authors \cite{mizok95,mizok04,wakis08}. A much smaller
value 2.5 eV (and $J_H=$ 0.25 eV) was used in the eleven-band $d$-$p$
model \cite{yamak10} (however this model is different as it includes
3D component --- namely the influence of neighbouring Na ions in
Na$_{0.5}$CoO$_2$ on CoO$_2$ layer). Other values suggested for these
systems are: 5.0 eV \cite{zhang04}; 4.5 eV \cite{kroll06} (from fits to
XAS experiments) and 4-8 eV used in LDA+$U$ approaches
\cite{wu05,zou04,lda1,lda2}. A much smaller value of 1.86 eV, being
renormalised by a factor of three from the atomic value due to strong
screening effects (similarly like it happens in cuprates) is reported
by Bourgeois {\it et al} \cite{bourg09}. Quite surprisingly, another
reference based on {\it ab initio} reports $U \simeq 4.1-4.8$ eV
\cite{land06}, i.e., larger values that those one could expect when
making comparison with very strong $J_H$ renormalisation reported in
the same paper \cite{land06}. Here we adopt a value $U=4.5$ eV, i.e.,
our ratio is $U/t_0$=15  (it is presumably large enough for
a strongly correlated cobalt oxide).

\section{Computational details}

We performed extensive computations for $6\times 6$ clusters with
periodic boundary conditions (PBC) to establish the ground state at
different doping
(at zero temperature $T=0$). Our reference (undoped) state with
Co$^{3+}$ ions contains $6\times 36$ electrons. For each doping, i.e.,
for fixed number of holes $n_h$ (number of deficient electrons) we
studied separately systems with different numbers of $n_{h\uparrow}$
--- up deficient electrons and $n_{h\downarrow}$ --- down deficient
electrons upon constraint $n_h=n_{h\uparrow}+n_{h\downarrow}$.
For nonmagnetic (or AF) systems with zero total magnetization we have
$n_{h\uparrow}=n_{h\downarrow}=\frac12 n_h$. All other possibilities
correspond to systems with non-zero total magnetization. Purely FM
states have $n_{h\uparrow}=0$ and $n_{h\downarrow}=n_h$. After long
screening of preliminary data we have established that the ground 
states are unpolarized, with $n_{h\uparrow}=n_{h\downarrow}=\frac12 n_h$.

Coming back to computations, there were two distinct steps made for
each parameter set and electron concentration:
($i$) first, the calculations within the single-determinant HF
approximation were performed, and
($ii$) in the next step the HF wave function was modified to include
the electron correlations by employing the local ansatz \cite{Sto80}.
This ansatz was successfully applied to several systems, {\it inter alia} 
to cuprates \cite{Ole87}, nickelates \cite{ni}, manganites \cite{bilayer}
and chemical bonds in molecular systems \cite{Ole86}. Here
the HF computations were run starting from each one of many different
initial conditions (to get unbiased results we considered up to 10000
nonhomogeneous random charge and spin
arrangements for each doping level). In addition, the symmetric
patterns known from the literature \cite{mizok04} were also included
as possible HF initial conditions and compared with the results
obtained for other configurations. For each one of starting initial
conditions we obtain on convergence a new HF wave function
$\vert\Psi_{\rm HF}\rangle$ which needs to be considered further to
implement local Coulomb correlations.
Thus, after completing the HF computations we performed correlation
computations to obtain the total energy and to identify the optimal
ground state configuration. Namely, the HF wave function
$|\Phi_0\rangle$ was modified to include the electron correlation
effects by using exponential local ansatz \cite{Sto80},
\begin{equation}
|\Psi\rangle = \exp\Big(-\sum_m \eta_mO_m\Big)|\Phi_0\rangle,
\label{la}
\end{equation}
where $\{O_m\}$ are local correlation operators. The values of
variational parameters $\{\eta_m\}$ are found by
minimizing the total energy,
\begin{equation}
E_{\rm tot}=\frac{\langle\Psi|H|\Psi\rangle}{\langle\Psi|\Psi\rangle}.
\label{etot}
\end{equation}
Here for the local correlation operators we use 25 operators which
optimize the density-density correlations,
\begin{equation}
O_m = \sum_i \delta n_{i\mu \uparrow}\delta n_{i\nu {\downarrow}},
\label{om}
\end{equation}
i.e., we use all possible combinations
$\mu,\nu=xy,yz,zx,x^2-y^2,3z^2-r^2$ of orbital indices. The symbol
$\delta$ in $\delta n_{i\mu\sigma}$ indicates that {\it only that part}
of $n_{i\mu\sigma}$ operator is included which annihilates one electron
in an occupied single particle state belonging to the HF ground state
$|\Phi_0\rangle$, and creates an electron in one of the virtual empty
states. The above local operators $O_m$ correspond to the subselection
of presumably most important electron-pair excitations within the
{\it ab initio\/} configuration-interaction method (for details see
\cite{bilayer} and \cite{ni}).

After obtaining the total energy for a given starting condition, we
repeat all the procedure from the beginning, i.e., we take the second,
third, fourth, ...  set of HF initial conditions and repeat all
computations to obtain the second, third, fourth, ... ,{\it etcetera},
candidate for a ground state wave function. The resulting set of total
energies was inspected and the few lowest ones were identified as
probable candidates for the true ground state. At this stage we
inspected the resulting charge and spin order (within the set of the
selected candidates) and prepared the second much smaller set of
initial HF conditions which on one hand were very similar to our
candidates and on the other hand we made small changes to enhance
local symmetry according to physical insights. The same procedure of
performing HF computations and adding local correlations was repeated
and the state with lowest energy was picked as our true ground state.
We emphasize that altogether such a procedure is very time and labour
consuming but it gives relatively high confidence that, {\it what we
identified as the ground state is indeed realized}, within the present 
effective model for the considered parameters and doping. We remark that 
the correlation contributions to the the total energy were found to be 
important for the correct identification of the ground state (as expected).

\section{Numerical results}

The first computational scan we performed and presented here is for the
ground state in undoped substance for varying crystal-field splitting 
$D_2$, see figure 1. We take a standard set of parameters as described in
section 2, which corresponds to a strongly correlated system (all in eV): 
\begin{equation}
t_0=0.3, \hskip .3cm t_1=0.05, \hskip .3cm U=4.5, \hskip .3cm D_1=0.105 
\label{para}
\end{equation}
and adopt the constrait $C=4B$
(the other parameters for the Jahn-Teller terms are given in the caption 
of figure 1). For a fixed value of Hund's exchange given by $B=0.1$ eV
and for $D_2<0.87$ eV we found the ground state to be 
totally charge-homogeneous with AF-like arrangement of high spins (close 
to $S=2$) and the electron configuration $t_{2g}^4e_g^2$ at each site; 
in a different systems such high spin states for Co$^{3+}$ ions are the 
subject of current interest \cite{Seo12,Sbo09}). 
After crossing the value of $D_2=0.88$ eV the ground state (again with
homogeneous charge distribution) becomes nonmagnetic (with spins $S=0$) 
as expected, and the electron configuration is $t_{2g}^6$. We performed 
additional computations within in the range $0.87\leq D_2\leq 0.88$ eV 
and found that the change of the spin state (and of magnetic order) 
order occurs abruptly (and it resembles a phase transition). Note that 
the bulk of computations performed in this paper was done for $D_2=1.25$ 
eV, i.e., well inside low spin (nonmagnetic) regime for cobalt-oxide 
compounds. Yet, finite doping generates high magnetic moments, see below.

\begin{figure}[t!]
\begin{center}
\includegraphics[width=8cm]{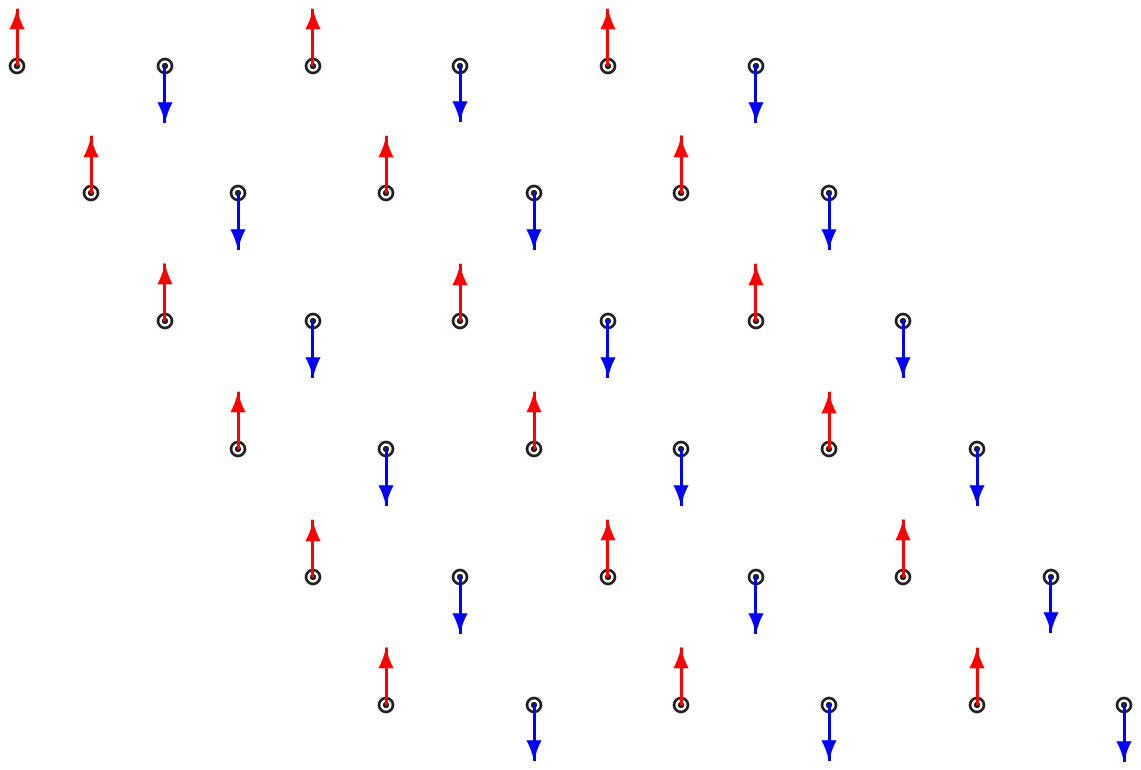}
\end{center}
\vskip 0.1cm
\begin{center}
\includegraphics[width=8cm]{cross.eps}
\end{center}
\caption{
{\it Top panel}--- High spin ground state with AF N\'eel order for zero 
doping for the triangular lattice obtained on a $6\times 6$ cluster
with PBC for $D_2=0.87$ eV and $B=0.1$ eV. Dots correspond to lattice 
sites and arrows indicate local high spin states (very close to
$S=2$). Note that at $D_2=0.88$ eV a drastic crossover takes place (for 
this value of $B$) and all spins collapse to zero when the $e_g$ states 
become virtually unoccupied (the electron occupation of each $e_g$ level 
is then $\approx$ 0.07).
{\it Lower panel}--- Phase diagram obtained by varying crystal-field 
splitting $D_2$ and Racah parameter $B$, for the undoped CoO$_2$ 
triangular plane with Co$^{3+}$ ions.
AF denotes a region of AF order with high spin ($S=2$) states of 
Co$^{3+}$ ions, and PA stands for nonmagnetic ground state ($S=0$) with 
empty $e_g$ orbitals. Diamonds are computational results, while the line 
is a guide for an eye. Other parameters (for both panels) as in equation 
\eref{para} and: $g_{\rm JT}=1.6$ eV\AA$^{-1}$, $K_{\rm JT}=13$ 
eV\AA$^{-2}$, $K_{\rm br}/K_{\rm JT}\gg 1$.
}
\end{figure}

Knowing that Co$^{3+}$ ions in the bulk system are in nonmagnetic $t_{2g}^6$
configuration, the other computations are done well inside the low spin
($S=0$) regime. Therefore we have used the fixed value of $D_2=1.25$ eV.
In spite of this rather high value of $D_2$ all the computations clearly
show that upon doping localized holes with high spin (close to $S=5/2$)
and with occupied $e_g$ levels appear as a generic feature of the ground
state. This is observed for any doping in the investigated doping range 
$0<x<1.0$. We have verified that these high spin states occur in the 
states characterized by very similar energies per doped hole,
\begin{equation}
E_h=\frac{1}{n_h}\left\{E_{\rm tot}(n_h)-E_{\rm tot}(n_h=0)\right\},
\end{equation}
where $n_h$ is the number of holes in the considered $6\times 6$
cluster. Energy per one doped hole $E_h$ increases almost in a linear 
way with increasing doping level, see figure 2. We remark that for each 
doping considered here we did not observe any significant Jahn-Teller 
static distortions associated with holes (they turn out to be small).

\begin{figure}[t!]
\centerline{
\includegraphics[width=8cm]{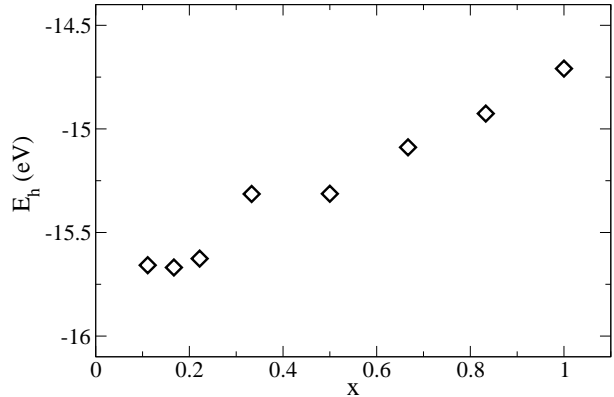}
}
\caption{%
Energy $E_h$ (HF + correlations) per doped hole for increasing doping 
$x$ in a $6\times 6$ cluster with PBC. For the reference system with no
holes (at $x=0$), the number of electrons is $6\times 36$,
the HF energy of the whole cluster is $E_{\rm HF}$ = 1400.538 eV and
the total energy (HF+correlations) is $E_{\rm tot}$ = 1400.002 eV.
Parameters as in equation \eref{para} and: 
$B=0.1$ eV, $D_2=1.25$ eV, $g_{\rm JT}=1.6$ eV\AA$^{-1}$, 
$K_{\rm JT}=13$ eV\AA$^{-2}$, $K_{\rm br}/K_{\rm JT}\gg 1$.
}
\end{figure}

We have not found any interesting local effects in the charge
distribution in the dilute limit when the system is doped by $n_h=1$ or 
2 holes within the $6\times 6$ cluster. In these cases the extra charge 
is distributed almost uniformly over the cluster atoms and all atoms are 
nonmagnetic. But already for a somewhat higher hole number $n_h=4$, 
corresponding to low doping $x=\frac{1}{9}$, two polaronic states are 
found, see figure 3(a). At each atom of the polaron the electron density 
is close to 5.2 and a high spin $S\simeq\frac{5}{2}$ arises. Large 
spins arise pairwise and are oriented in the opposite way --- we suggest
that they would give a nonmagnetic singlet state when the quantum spin
fluctuations were also included. For this low level of doping the 
polarons are isolated and no phase separation is found.

\begin{figure}[t!]
\centerline{(a)
\includegraphics[width=8cm]{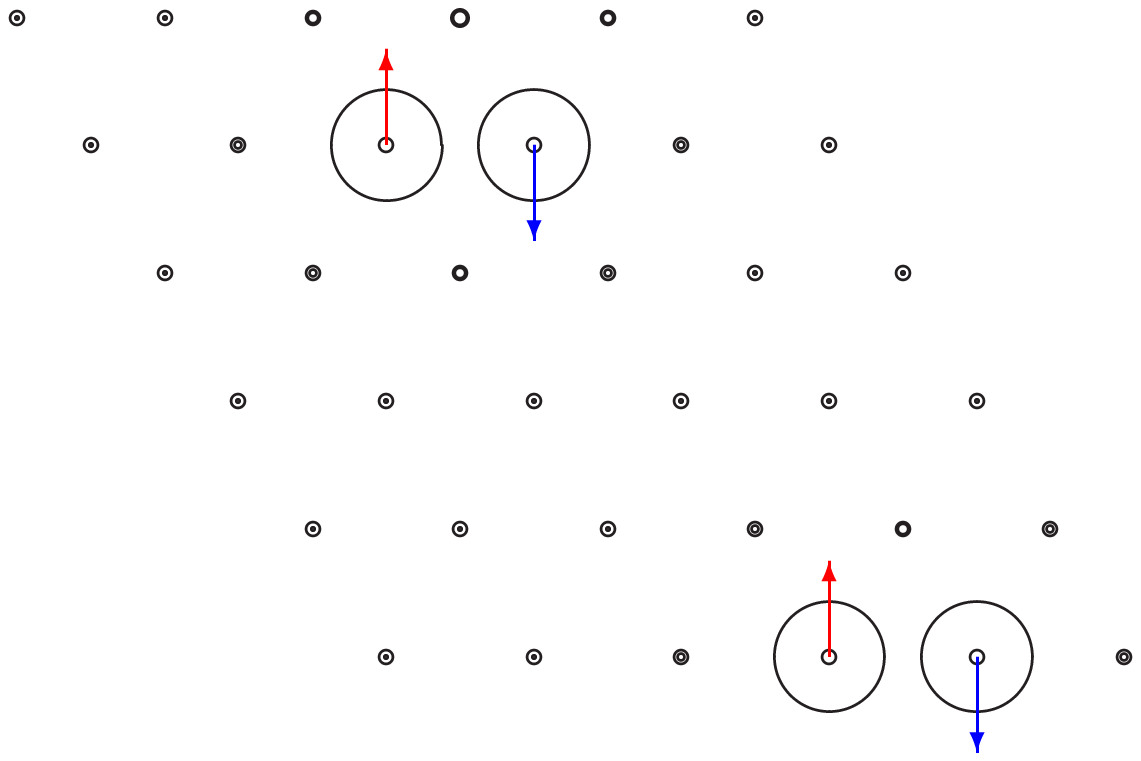}}
\vskip 0.5cm
\centerline{(b)
\includegraphics[width=8cm]{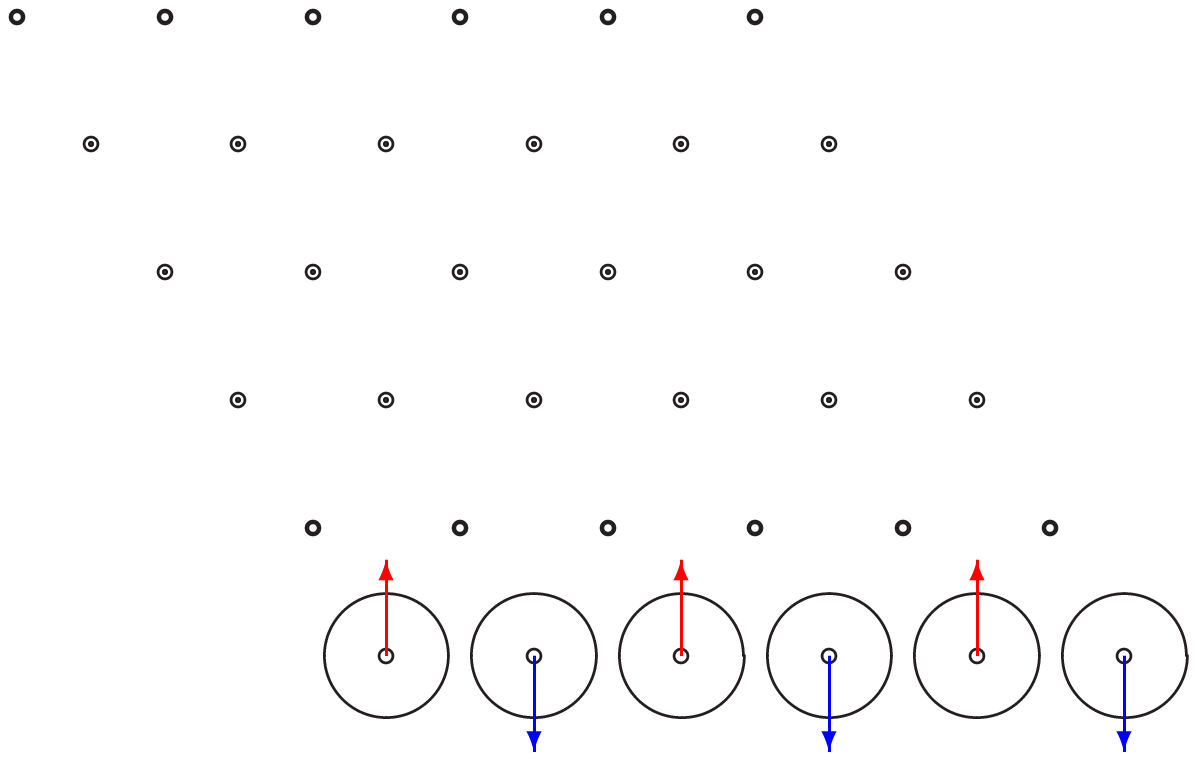}}
\vskip 0.5 cm
\centerline{(c)
\includegraphics[width=8cm]{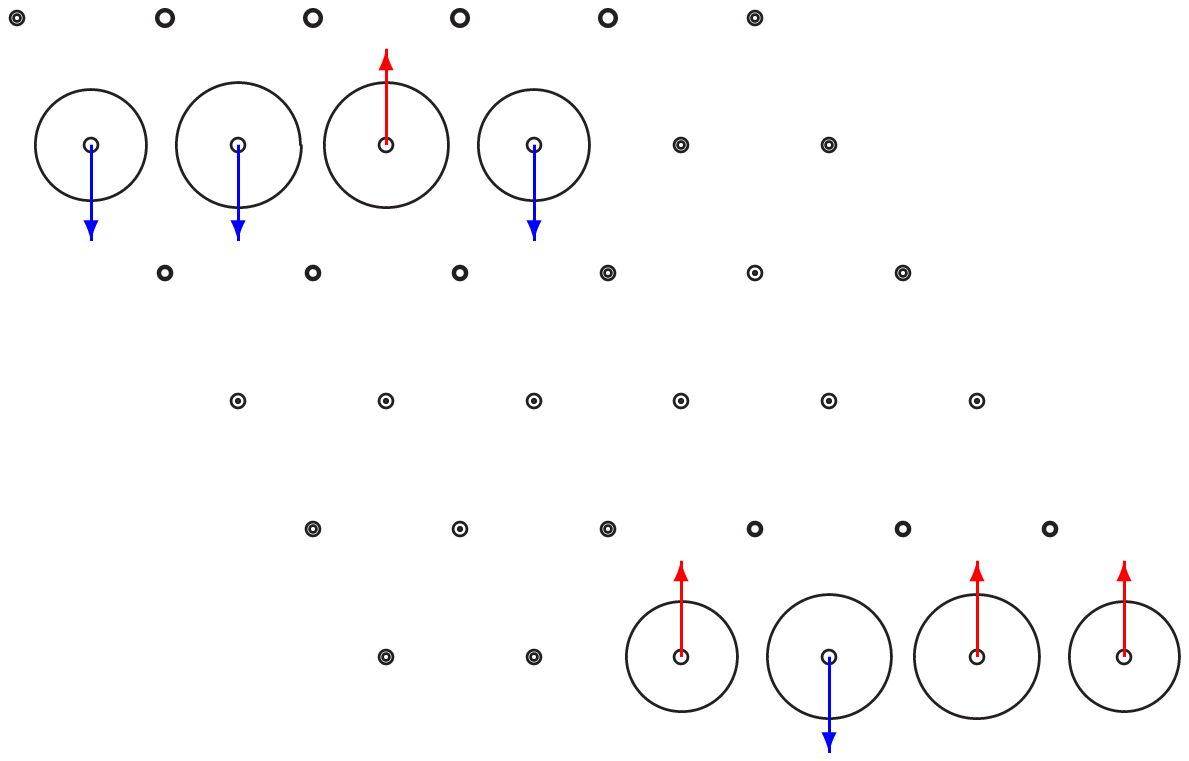}}
\caption{High spin states at doped holes (spin values very close to 
$S=\frac{5}{2}$) in ground states obtained in the lowdoping regime: 
(a) $x=\frac{1}{9}$ (4 doped holes,  upper panel), 
(b) $x=\frac{1}{6}$ (6 doped holes, middle panel), and 
(c) $x=\frac{2}{9}$ (8 doped holes,  lower panel). 
Dots correspond to lattice sites and arrows indicate high spin states.
Circle are corresponds to hole charges (each big circle denotes a hole 
with $\sim 0.8e$ missing). 
Parameters as in equation \eref{para} and: 
$B=0.1$ eV, $D_2=1.25$ eV, $g_{\rm JT}=1.6$ eV\AA$^{-1}$, 
$K_{\rm JT}=13$ eV\AA$^{-2}$, $K_{\rm br}/K_{\rm JT}\gg 1$;
we remind that the $t_{2g}-e_g$ splitting is well inside low spin 
regime of an undoped CoO$_2$ plane for the present value of $D_2$.
}
\end{figure}

For low doping $x=\frac{1}{6}$ and $x=\frac{2}{9}$, see figures 3(b) and
3(c), we observe again high spin states of doped holes, essentially the 
same spin values $S\simeq\frac{5}{2}$ as for the case of $x=\frac{1}{9}$.
The holes with low or intermediate spin values are absent in all cases
(but they are found in metastable states, i.e., for local HF minima
with higher energies). One finds tendency to form locally bound singlet
states, i.e., up-spin and down-spin pairs with both spins placed close 
to each other. For $x=\frac{1}{6}$ the stripe-like one-dimensional (1D)
structure (in each sixth line with AF order) is clearly emerging, see 
figure 3(c). This is a precursor state of the ordered 1D structures 
which occur at higher hole doping, see below.

\begin{figure}[t!]
\centerline{(a)
\includegraphics[width=8cm]{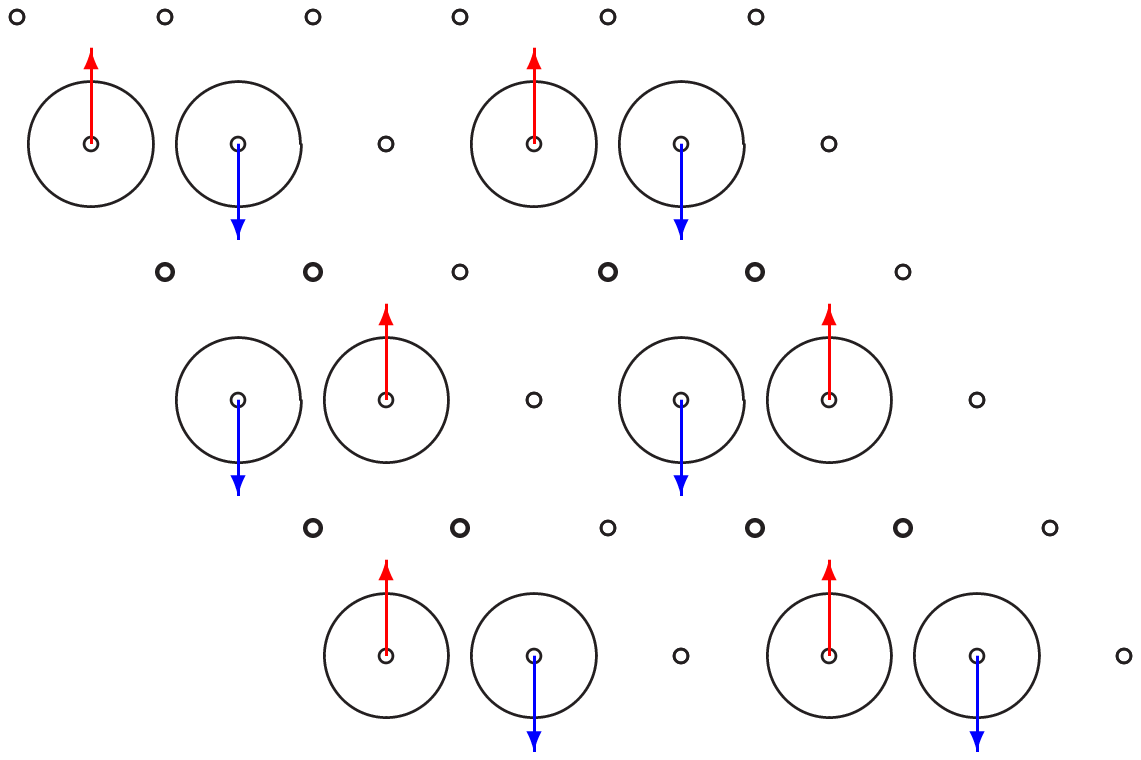} }
\vskip 0.5cm
\centerline{(b)
\includegraphics[width=8cm]{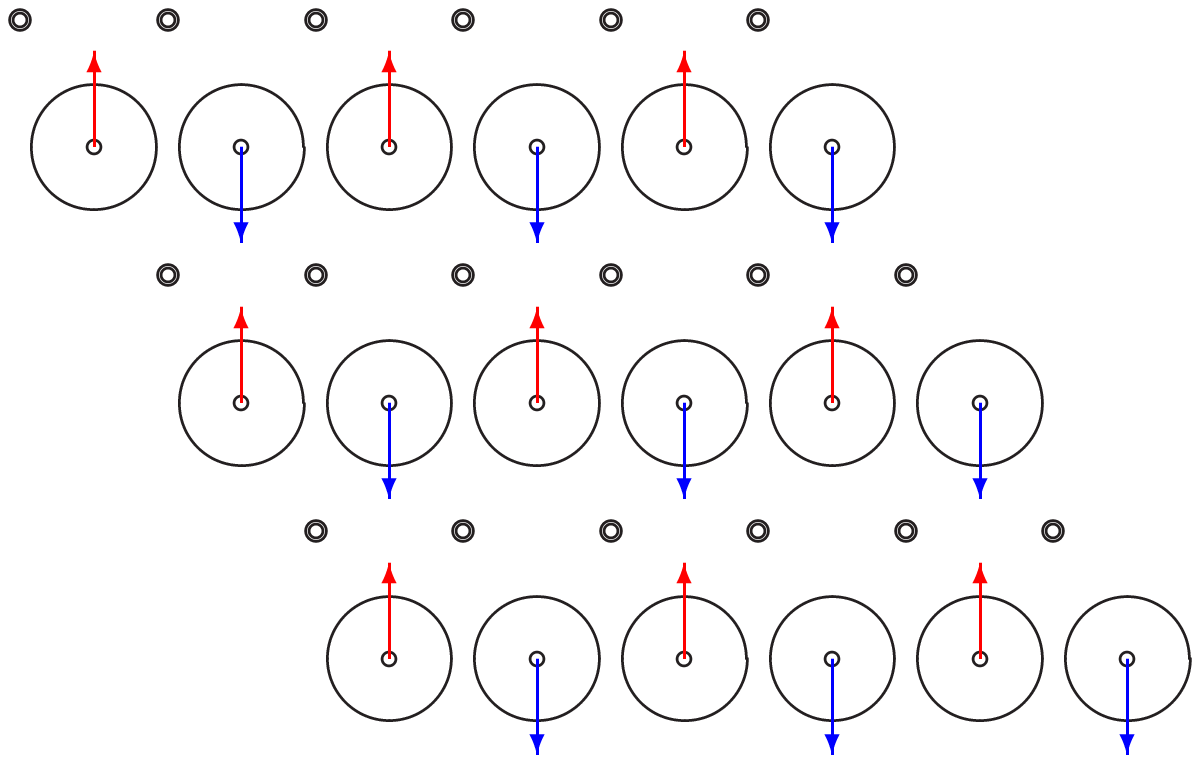}}
\caption{Stripe-like structures of high spin doped holes (spin values
at sites with high hole density (circles) are very close to
$S=\frac{5}{2}$) in ground states for dopings: 
(a) $x=\frac{1}{3}$ (upper panel) and 
(b) $x=\frac{1}{2}$ (lower panel). 
Dots correspond to lattice sites, arrows indicate high spin states and 
each circle corresponds to $\sim 0.8e$ hole charge.
Parameters as in equation \eref{para} and: 
$B=0.1$ eV, $D_2=1.25$ eV, $g_{\rm JT}=1.6$ eV\AA$^{-1}$, 
$K_{\rm JT}=13$ eV\AA$^{-2}$, $K_{\rm br}/K_{\rm JT}\gg 1$.
High HOMO-LUMO gaps indicate that the ground states are insulating.
}
\end{figure}

At this place we would like to make a somewhat obvious but still very
important observation that the doping level in the cluster $x$, i.e.,
the number of deficient electrons per site ($x=n_h/N$, where $N=36$)
and the subscript $x$, say in the chemical formula Na$_{1-x}$CoO$_2$,
are not the same. For low doping levels in various transition metal
oxides they may turn out to be approximately the same but there is no
such guarantee for cobaltates. The well known example are
YBa$_2$Cu$_3$O$_{6+x}$ superconductors, where the actual hole
concentration $x$ is quite distinct from the chemical doping
\cite{zaa88}. One should keep this observation in mind
when trying to compare any computational results reported here with
the experimental data for particular cobalt oxides.

\begin{figure}[t!]
\centerline{(a)
\includegraphics[width=8cm]{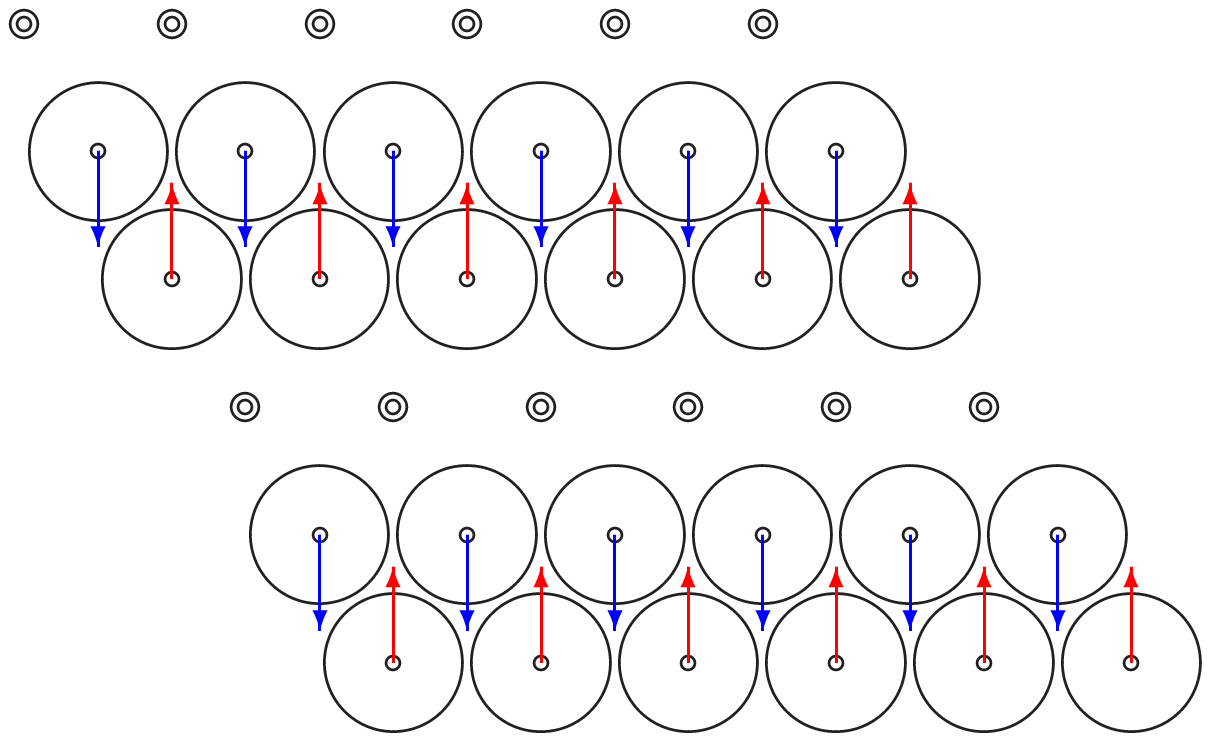} }
\vskip 0.5cm
\centerline{(b)
\includegraphics[width=8cm]{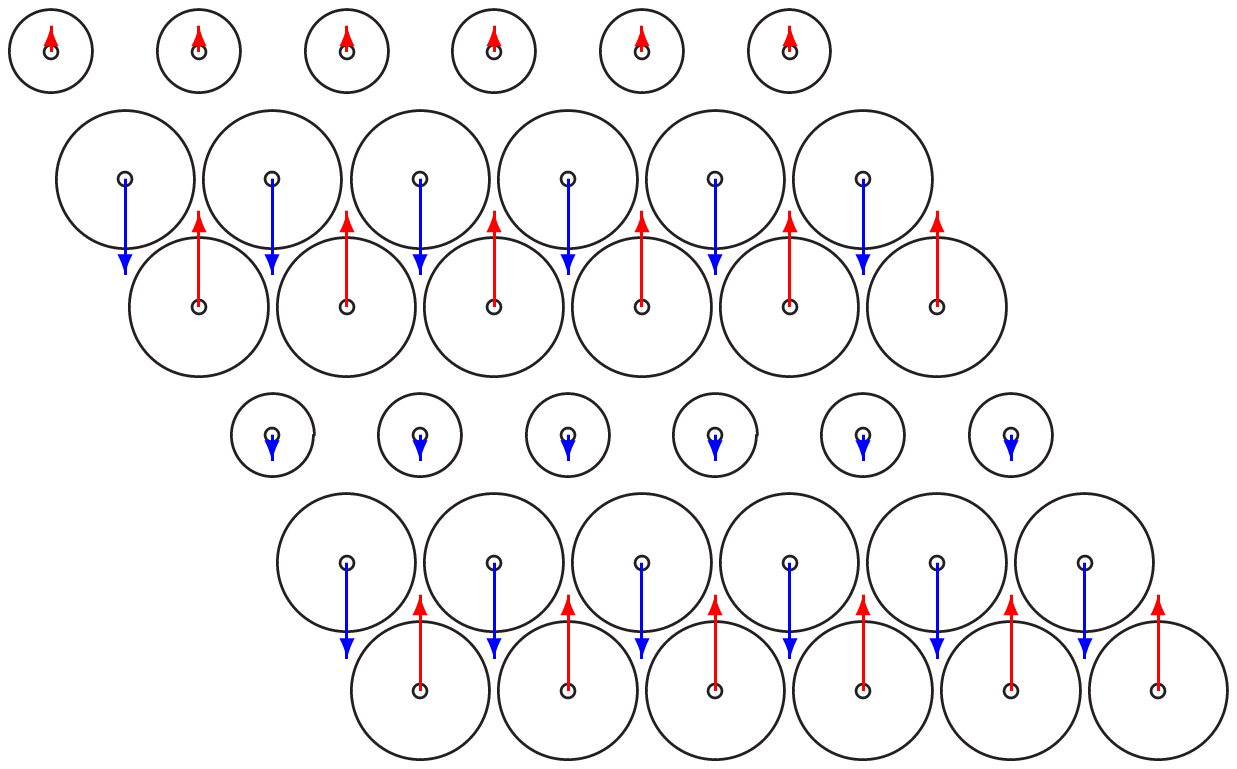} }
\vskip 0.5cm
\centerline{(c)
\includegraphics[width=8cm]{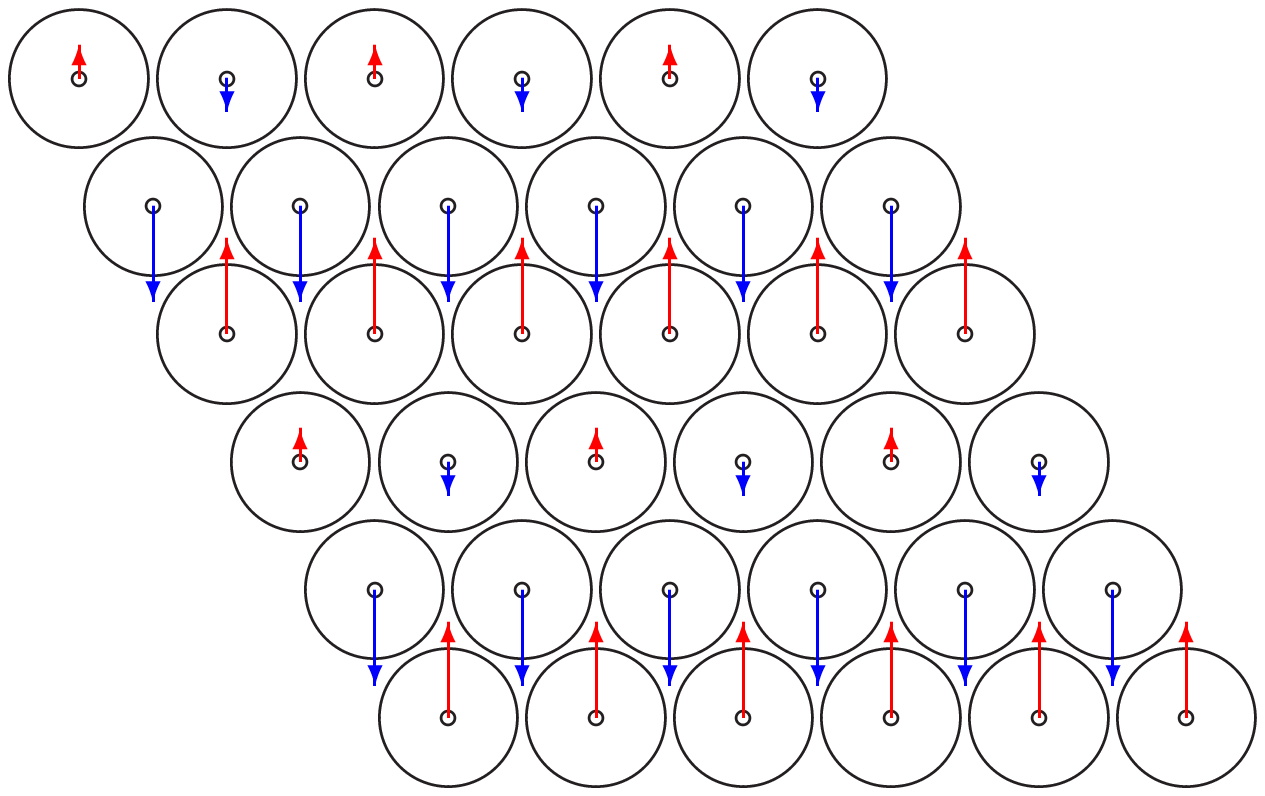} }
\caption{Thick stripe-like walls of high spin doped holes (spin values
are very close to $S=\frac{5}{2}$) in ground states obtained for doping:
(a) $x =\frac{4}{6}$ (upper panel),
(b) $x =\frac{5}{6}$ (middle panel), and
(c) $x =$1 (lower panel).
Parameters as in equation \eref{para} and: 
$B=0.1$ eV, $D_2=1.25$ eV, $g_{\rm JT}=1.6$ eV\AA$^{-1}$, 
$K_{\rm JT}=13$ eV\AA$^{-2}$, $K_{\rm br}/K_{\rm JT}\gg 1$.
Hole charges are $\sim 1e$ at high spin sites, while smaller circles 
correspond to $\sim 0.6 e$ charge. Low spin states are found at higher 
doping: 
(b) $S\sim\frac{1}{4}$, and 
(c) $S\sim \frac{1}{2}$.
Relatively high HOMO-LUMO gaps indicate that the ground states are insulating.
}
\end{figure}

Consider now doping increasing further beyond $x=\frac{2}{9}$. One finds 
then an interesting evolution of polaronic structures which self-organize. 
High-spin states arise again and the number of ions with
$S\simeq\frac{5}{2}$ spins is equal to the number of doped holes, while
low spin and intermediate spin states are absent. There is a pronounced 
tendency to form first hole pairs with singlet-like spin states and to 
place such singlets maximizing the distance one from another. At doping 
of $x=\frac{1}{3}$ pairs of polarons are ordered indeed in a pattern 
which maximizes their distances from one another, see figure 4(a). 
At half-doped system ($x=\frac12$) the lines of polarons form instead 
and the spin order along each line is AF, see figure 4(b). This result 
corroborates with the insulating character of both states which makes 
superexchange between pairs of Co$^{4+}$ ions at ions doped with holes 
the most important magnetic exchange process.

Finally, a highly doped regime $\frac12<x<1$ is characterized by 
stripe-like ground states, with lines of weakly doped sites of (almost) 
nonmagnetic ions in between the ordered lines of polarons, see figures 
5(a) and 5(b). Having the $6\times 6$ cluster size, we study the dopings 
where the number of holes $n_h$ is divisible by 3. The holes occupy 
predominantly four lines in the considered clusters at $x=\frac{4}{6}$ 
and $x=\frac{5}{6}$, and the sites doped by one hole to Co$^{4+}$ ionic 
configurations have high spin $S\simeq\frac{5}{2}$. 
As the electron density is there somewhat higher than 5 electrons per 
site, the magnetic moments order along the lines to the FM state which 
reflects weak double exchange mechanism \cite{Ole11} in cobaltates. For 
other dopings (not shown) this high symmetry of the solutions is lifted
--- however, the same trends (as visible in figures 3, 4 and 5) to form
symmetric ground states for ``magic dopings'' is still clearly visible.

It is remarkable that such stripe-like structure survives even in the
limit of $x=1$, see figure 5(c), and was obtained from unbiased initial
configurations as the most stable state. Surprisingly, the lines which 
separate the ordered structure of high spin polarons have here low spins
$S\simeq\frac12$, and these spins have AF order along their lines, 
in contrast to the FM order along lines of large $S\simeq\frac{5}{2}$ 
spins. Such a state may be understood as following from weak AF 
superexchange which becomes active when one $t_{2g}$ hole couples the 
neighbouring low spin ions. At the same time, absence of high spins at 
every third line reduces frustration of magnetic interactions.

\section{Discussion and Summary}

The results we obtained for increasing doping in CoO$_2$ planes are
somewhat unexpected. A systematic trend was found that doping creates
high spin ($S=\frac{5}{2}$) states and holes are pretty well localized 
on Co$^{4+}$ ions. Note that orbital degrees of freedom are saturated 
here (for $S=\frac{5}{2}$ states) so Jahn-Teller distortions or 
spin-orbital entanglement are not expected. So far, there is no clear 
and direct experimental support for such hole localization in form of 
high spin ($S=\frac{5}{2}$) states, and they were not reported in 
doped Na-compounds. The only exception is found in layered
Li$_{1-x}$CoO$_2$ for very low doping level $0<x<0.06$ \cite{her08}.
Also for $x=0.5$ neutron scattering study of Na$_{0.5}$CoO$_2$ gives
the charge and spin order in agreement with that shown in figure 4(b), 
but the authors \cite{gas06} interpret their data in terms of low 
$S=\frac12$ spins. For lower Na concentration $1-x\sim 0.3$ there is 
some evidence of antiferromagnetic spin-spin correlations \cite{ohi10}.

The central result concerning the electronic structure, is that all 
the investigated ground states are insulating (sizable HOMO-LUMO gaps). 
On the contrary, it is well known that the ground state of 
Na$_{0.75}$CoO$_2$ is metallic \cite{garba08}. 
Other systems where high spins \cite{hollm08,merz11} or stripe 
structures \cite{cwik09} were found are clearly beyond the assumptions
made within the present model (due to several reasons; to give
one simple example, due to the presence of Co$^{2+}$ ions). This in our 
opinion does not invalidate the results we obtained upon assumption that 
the system is a truly regular 2D triangular plane. Namely, the
experimental data \cite{yokoi05} on Na$_{1-x}$CoO$_2$ and subsequent
theoretical investigation \cite{yamak10,hel06} clearly show that the
assumption made about 2D uncoupled layers in Na$_x$CoO$_2$ is an 
idealization that does not reflect the properties of real systems. 
Only for smaller content of Na ions (higher doping $x$) the 2D nature of
electrons in Na$_{1-x}$CoO$_2$ is enhanced (and the antiferromagnetic
spin correlation increases) \cite{yok05}. Thus, there is a significant
3D component to the real layered system and therefore 2D computations
are expected to show rather different charge and magnetization
distribution from the experimental data.

At present, our results can be treated as a prediction pending
until some new truly 2D system with a triangular lattice is discovered
and investigated. Also for the already discussed systems with low hole
concentration $x$ one might expect that entangled spin states with up
and down high spin pairs would form. Observation of such elangled states 
in real systems, their treatment in the theory, and search for 
intermediate spin ($S=1$) states which could be stabilized by quantum 
effects beyond the present theory \cite{Ole12}, provide experimental 
challenges in the physics of cobal oxides. 

Summarizing, we have established a generic trend that doping of CoO$_2$
planes induces localized hole states with high $S=\frac{5}{2}$ spins. 
This releasing of spin states at the doped Co$^{4+}$ ions with reduced 
electron density and partly filled $e_g$ orbitals makes it necessary to 
consider the full five-band model including Co($3d$) orbitals 
\cite{Kha08}, in spite of having (almost) empty $e_g$ orbitals in undoped 
compounds with a higher electron density. The present study suggests 
as well that one has to use then the full Hund's exchange tensor for a 
realistic description. The doped holes first {\it self-organize}
into polaronic states consisting of two holes each in the regime of low
doping, and next form ordered 1D structures when doping approaches
$x=\frac13$. We suggest that superexchange between $S=\frac{5}{2}$ spins
is the dominating magnetic interaction which is responsible for the
antiferromagnetic spin order along the 1D lines up to half-doping
($x=\frac12$). Higher doping generates triangles occupied by three spins
$S=\frac{5}{2}$ spins and superexchange interactions are frustrated. In 
this regime the system selects antiferromagnetic order with lines of
ferromagnetic spins which are believed to follow from weak FM double
exchange mechanism \cite{Ole11}. Surprisingly, low spin $S=\frac12$
states survive {\it even} in the fully doped case ($x=1$) and serve to
stabilize the magnetic order of large spins along the AF 1D structures
with pairs of lines containing ferromagnetic spins each. One could
expect however that such an ordered phase will be destabilized by 
quantum fluctuations and a disordered magnetic state would arise instead.

As a final remark let us address the question of possible future
extensions of this model approach: Are possible changes of the
Hamiltonian parameters not needed to describe Na$_{1-x}$CoO$_2$ 
and how robust the results might be with respect to them?
It would be indeed quite interesting to repeat all the computations for
different sets of the Hamiltonian parameters. However, as mentioned
above, such computations are very time consuming so further numerical
studies could be motivated only by experimental information
concerning more precise values of the parameters for the systems with
CoO$_2$ planes. We expect that this information will become available
due to future experiments.

\ack

We kindly acknowledge financial support by the Polish National Science 
Center (NCN) under Project No. 2012/04/A/ST3/00331.

\section*{Appendix: Hopping elements in the effective model}

Here we present supplementary data which justify the choice of hopping
parameters and give more details on the values of hopping elements which 
are used in the kinetic energy \eref{Hkin} in the effective model for 
$3d$ electrons \eref{model} in section 2, see tables 1 and 2. We begin 
with the hopping elements resulting from indirect Co-O-Co hopping, and 
next present direct Co-Co hopping. These two different sets of hopping 
elements are given by two parameters, $t_0$ and $t_1$, respectively. 

\begin{table}[t!]
\caption{Effective hopping elements $t_{i\mu,j\nu}$ between orbitals 
$\mu$ and $\nu$ at sites $i$ and $j$ resulting 
from indirect cobalt-oxygen-cobalt transitions in a triangular lattice 
as obtained using Slater-Koster rules \cite{livermore} and perturbation 
theory \cite{Zaa93}, in units of $t_0 = P^2_{pd\pi}/\Delta $. Bond 
directions are given by lattice vectors $\mathbf{a}_n$, see equation 
\eref{vectors}. Furthermore, we use the ratio $P_{pd\sigma}/P_{pd\pi}=-2.0$ 
\cite{mizok95,wakis08,mizok04}. The entries for $\nu<\mu$ were omitted 
as symmetry implies that $t_{i\mu,j\nu}=t_{i\nu,j\mu}$.
}
\begin{center}
\begin{tabular}{ccccc} \hline
 $i$ & $j$  &  $\mu$  & $\nu$ &   $t_{i\mu, j\nu}$   \\ \hline
{\bf{0}}&   {\bf{a}}$_1$& $ xy $ & $ zx$& $t_0$  \\
{\bf{0}}&   {\bf{a}}$_1$& $ yz $ & $ 3z^2-r^2$ & $ -t_0$  \\
{\bf{0}}&   {\bf{a}}$_1$& $ yz $ & $ x^2-y^2$ & $ \sqrt{3} t_0$  \\   \hline
{\bf{0}}&   {\bf{a}}$_2$& $ xy $ & $yz$ & $t_0$\\
{\bf{0}}&   {\bf{a}}$_2$& $ zx $ & $ 3z^2-r^2 $ & $ -t_0$\\
{\bf{0}}&   {\bf{a}}$_2$& $ zx $ & $ x^2-y^2 $ & $ -\sqrt{3} t_0$\\   \hline
{\bf{0}}&   {\bf{a}}$_3$& $ yz $ & $ zx$ & $t_0$\\
{\bf{0}}&   {\bf{a}}$_3$& $ xy $ & $3z^2-r^2 $& $  2 t_0$ \\  \hline
\end{tabular}
\end{center}
\end{table}

\begin{table}[t!]
\begin{center}
\caption{The direct cobalt-cobalt hopping elements $t_{i\mu,j\nu}$ 
between orbitals $\mu$ and $\nu$ at sites $i$ and $j$ for the triangular 
lattice, in units of $t_1\equiv\frac{1}{2}P_{dd\pi}$, and with 
$P_{dd\sigma}/P_{dd\pi}=-2.0$ \cite{wakis08}. The symmetry implies that 
entries for $\nu<\mu$ are the same as those for $\nu>\mu$.}
\begin{tabular}{ccccc} \hline
 $i$ & $j$  &  $\mu$  & $\nu$ &   $t_{i\mu, j\nu}$   \\ \hline
{\bf{0}}&   {\bf{a}}$_1$& $xy $&$ xy$ & $ t_1$     \\
{\bf{0}}&   {\bf{a}}$_1$& $ xy $&$zx$ & $ -t_1$     \\
{\bf{0}}&   {\bf{a}}$_1$& $ yz $&$yz$ & $ -3 t_1$     \\
{\bf{0}}&   {\bf{a}}$_1$& $ yz $&$x^2-y^2$ & $ -\frac{3}{2} t_1$     \\
{\bf{0}}&   {\bf{a}}$_1$& $ yz $&$3z^2-r^2$ & $  \frac{\sqrt{3}}{2} t_1$     \\
{\bf{0}}&   {\bf{a}}$_1$& $ zx $&$zx$ & $  t_1$     \\
{\bf{0}}&   {\bf{a}}$_1$& $ x^2-y^2 $&$x^2-y^2$ & $ -\frac{1}{4} t_1$     \\
{\bf{0}}&   {\bf{a}}$_1$& $ x^2-y^2 $&$3z^2-r^2$ & $  \frac{3\sqrt{3}}{4} t_1$     \\
{\bf{0}}&   {\bf{a}}$_1$& $ 3z^2-r^2 $&$3z^2-r^2$ & $ \frac{5}{4} t_1$     \\   \hline
{\bf{0}}&   {\bf{a}}$_2$& $ xy $&$xy$ & $  t_1$     \\
{\bf{0}}&   {\bf{a}}$_2$& $ xy $&$yz$ & $  -t_1$     \\
{\bf{0}}&   {\bf{a}}$_2$& $ yz $&$yz$ & $  t_1$     \\
{\bf{0}}&   {\bf{a}}$_2$& $ zx $&$zx$ & $  -3 t_1$     \\
{\bf{0}}&   {\bf{a}}$_2$& $ zx $&$x^2-y^2$ & $ \frac{3}{2} t_1$     \\
{\bf{0}}&   {\bf{a}}$_2$& $ zx $&$3z^2-r^2$ & $ \frac{\sqrt{3}}{2} t_1$     \\
{\bf{0}}&   {\bf{a}}$_2$& $ x^2-y^2 $&$x^2-y^2$ & $ -\frac{1}{4} t_1$     \\
{\bf{0}}&   {\bf{a}}$_2$& $ x^2-y^2 $&$3z^2-r^2$ & $ -\frac{3\sqrt{3}}{4} t_1$     \\
{\bf{0}}&   {\bf{a}}$_2$& $ 3z^2-r^2 $&$3z^2-r^2$ & $ \frac{5}{4} t_1$     \\   \hline
{\bf{0}}&   {\bf{a}}$_3$& $ xy $&$xy$ & $ -3 t_1$     \\
{\bf{0}}&   {\bf{a}}$_3$& $ xy $&$3z^2-r^2$ & $ -\sqrt{3} t_1$     \\
{\bf{0}}&   {\bf{a}}$_3$& $ yz $&$yz$ & $   t_1$     \\
{\bf{0}}&   {\bf{a}}$_3$& $ yz $&$zx$ & $   -t_1$     \\
{\bf{0}}&   {\bf{a}}$_3$& $ zx $&$zx$ & $   t_1$     \\
{\bf{0}}&   {\bf{a}}$_3$& $ x^2-y^2 $&$x^2-y^2$ & $  2 t_1$     \\
{\bf{0}}&   {\bf{a}}$_3$& $ 3z^2-r^2 $&$3z^2-r^2$ & $   -t_1$     \\ \hline
\end{tabular}
\end{center}
\end{table}

The hybridization elements $P_{pd\sigma}$ and $P_{pd\pi}$ are the
appropriate Slater-Koster interatomic integrals \cite{livermore} and
$\Delta$ is charge-transfer energy between bare cobalt $3d$ level and
oxygen $2p$ level \cite{mizok95,koshi03,mizok04}.
The original estimates for $P_{pd\sigma}$, $\Delta$ and $t_0$ are,
respectively:
1.8 eV, 2.0 eV and 0.67 eV \cite{mizok95};
2.5 eV, 2.0 eV and 0.35 eV \cite{mizok04};
2.35 eV, 2.9 eV and 0.34 eV \cite{kroll06};
2.3 eV, 1.0 eV (here an effective value of  $\Delta$ is given) and
1.1 eV  \cite{wakis08};
1.4 eV, 3.2 eV and 0.15 eV  \cite{zou04}.
In general, the quoted values, if studied within multiband HF approaches,
differ from those coming out when {\it ab initio} results are combined
with (i.e., they are fitted to) particular experimental results.
This large variation of parameters occurs because in the first group of 
papers the correlations are not included (they could be included only in 
an {\it a posteriori} HF treatment). On the contrary, in the second 
group of papers effective models are constructed for the description
of particular experimental data and have the correlations included
(within the effective Hamiltonian parameters) just from the beginning.
On top of it effective models use the Hamiltonian parameters which come
out from complicated renormalisation and/or superposition of various
physical ingredients. The largest difference in the above mentioned two
groups of papers can be expected for the values of $\Delta$.
Finally, let us note that various {\it ab initio}-like evaluations and
other direct estimates of $t_0$ are: $\sim$ 0.1-0.3 eV \cite{land06}, 
0.1 eV \cite{khal,kroll,bourg07,bourg09} and (already mentioned) 0.15 eV
\cite{zou04}. Here we take $t_0=P^2_{pd\pi}/\Delta=0.3$ eV.

In addition, the ratio $P_{pd\sigma}/P_{pd\pi}$ has to be fixed. 
In a simplified approach $P_{pd\sigma}/P_{pd\pi}=-\sqrt{3}$, see 
\cite{harr05}; other reported values are higher: 
$P_{pd\sigma}/ P_{pd\pi}=-2.16$ \cite{mizok95,wakis08,mizok04} and 
$P_{pd\sigma}/P_{pd\pi}=-2.35$ \cite{kroll06}. Here we take 
$P_{pd\sigma}/ P_{pd\pi}=-2.0$ --- just for the sake of simplicity. 
With this latter choice the entries in table 1 representing hopping 
elements for pairs of different orbitals at neighbouring Co ions are 
simpler while the qualitative results of this study do not change when
a slightly different value of the ratio $P_{pd\sigma}/ P_{pd\pi}$ is 
chosen.

Unfortunately, much less is known about direct cobalt-cobalt hopping
elements --- the majority of authors assume that they are negligible. 
Here we adopt the ratio $P_{dd\sigma}/P_{dd\pi}=-2.0$ and take a value 
$P_{dd\pi}=0.1$ eV, following Bourgeois {\it et al} \cite{bourg07}. 
We also remark that according to Harrison rules 
$P_{dd\sigma}/P_{dd\pi}=-1.5$ \cite{harr05}, but having so small direct 
$d$-$d$ hopping elements no qualitative changes of the results are
expected when the above ratio would be taken instead. Complete list of
direct cobalt-cobalt hopping elements for different pairs of orbitals at 
nearest neighbour Co ions is given in table 2.


\section*{References}

\begin{thebibliography}{99}

\bibitem{Ima98} Imada M, Fujimori A and Tokura Y 1998
                   {\it Rev. Mod. Phys.} \textbf{70} 1039

\bibitem{Kor96} Korotin M A, Eshov Yu A, Solovyev I V, Anisimov V I,
                   Khomskii D I and Sawatzky G A 1996
                   {\it Phys. Rev.} B \textbf{54} 5309

\bibitem{Dag01} Dagotto E, Hotta T and Moreo A 2001
                   {\it Phys. Rep.} \textbf{344} 1 \\
                Dagotto E 2005
                   {\it New J. Phys.} \textbf{7} 67 \\ 
                Wei\ss{}e A and Fehske H 2004
                   {\it New J. Phys.} \textbf{6} 158 
                   
\bibitem{Kov10} Kovaleva N N, Ole\'s A M, Balbashov A M, Maljuk A,
                   Argyriou D N, Khaliullin G and Keimer B 2010                 
                   {\it Phys. Rev.} B \textbf{81} 235130

\bibitem{Fei99} Feiner L F and Ole\'s A M 1999
                   {\it Phys. Rev.} B \textbf{59} 3295  \\
                Ole\'s A M, Khaliullin G, Horsch P and Feiner L F 2005 
                   {\it Phys. Rev.} B \textbf{72} 214431 

\bibitem{Foo04} Foo M L, Wang Y Y, Watauchi S, Zandbergen H W, He T,
                   Cava R J and Ong N P 2004
                   {\it Phys. Rev. Lett.} \textbf{92} 247001

\bibitem{deV05} de Vaulx C, Julien M-H, Berthier C, Horvati\'c M, Bordet P, 
                   Simonet P V, Chen D P and Lin C T 2005
                   {\it Phys. Rev. Lett.} \textbf{95} 186405 \\
                Lang C, Bobroff J, Alloul H, Mendels P, Blanchard N 
                   and Collin G 2005
                   {\it Phys. Rev.} B \textbf{72} 094404 

\bibitem{Kha08} Khaliullin G and Chaloupka J 2008
                   {\it Phys. Rev.} B \textbf{77} 104532

\bibitem{bourg09} Bourgeois A, Aligia A A and Rozenberg M J 2009
                   {\it Phys. Rev. Lett.} \textbf{102} 066402

\bibitem{mizok04} Mizokawa T 2004
                   {\it New Journal of Physics} \textbf{6} 169

\bibitem{bourg07} Bourgeois A, Aligia A A, Kroll T and N\'u\~{n}ez-Regueiro M D 2007
                   {\it Phys. Rev.} B \textbf{75} 174518

\bibitem{kroll06} Kroll T, Aligia A A and Sawatzky G A 2006
                   {\it Phys. Rev.} B \textbf{74} 115124

\bibitem{koshi03} Koshibae W and Maekawa S 2003
                   {\it Phys. Rev. Lett.} \textbf{91} 257003

\bibitem{inder05} Indergand M, Yamashita Y, Kusunose H and Sigrist M 2005
                   {\it Phys. Rev.} B \textbf{71} 214414

\bibitem{yamak10} Yamakawa Y and Ono Y 2007
                   {\it J. Phys.: Condens. Matter} \textbf{19} 145289 \\
                Yamakawa Y, Watanabe N and Ono Y 2010
                   {\it J. Phys.: Conf. Series} \textbf{200} 012233

\bibitem{livermore} Slater C and Koster G F 1954
                   {\it Phys. Rev.} \textbf{94} 1498 \\
                Mehl M J
                   {\it Slater-Koster Tight-Binding Matrix Elements:}\\
                   http://cst-www.nrl.navy.mil/users/mehl/sk-param.html

\bibitem{Zaa93} Zaanen J and Ole\'s A M 1993
                   {\it Phys. Rev.} B \textbf{48} 7197

\bibitem{toyoz66} Toyozawa Y and Inoue M 1996
                   {\it J. Phys. Soc. Jpn.} \textbf{21} 1663

\bibitem{bilayer} Ro\'sciszewski K and Ole\'s A M 2008
                   {\it J. Phys.: Condens. Matter} \textbf{20} 365212 \\
                Ro\'sciszewski K and Ole\'s A M 2010
                   {\it J. Phys.: Condens. Matter} \textbf{22} 425601

\bibitem{gJT}   Pradheesh R, Nair H S, Sankaranarayanan V and Sethupathi K 2012
                   {\it Eur. J. Phys.} B \textbf{85} 260

\bibitem{kroll} Kroll T 2006
                   {\it On the structure of layered sodium cobalt oxides},
                   Ph.D. thesis, Technischen Universit\"at Dresden, Dresden

\bibitem{pilla08} Pillay D, Johannes M D, Mazin I I and Andersen O K 2008
                   {\it Phys. Rev.} B \textbf{78} 012501

\bibitem{zou04} Zou L-J, Wang J L and Zeng Z 2004
                   {\it Phys. Rev.} B \textbf{69} 132505

\bibitem{zaliz00} Zaliznyak I A, Hall J P, Tramquada J M, Erwin R
                   and Morimoto Y 2000
                   {\it Phys. Rev. Lett.} \textbf{85} 4353

\bibitem{huang04} Huang Q, Foo M L, Pascal R A Jr, Lynn J W, Toby B H,
                   He T, Zandbergen H W and Cava R J 2004
                   {\it Phys. Rev.} B \textbf{70} 184110

\bibitem{merz11} Merz M, Fuchs D, Assmann A, Uebe S, v.Loehneysen H,
                   Nagel P and Schuppler S 2011
                   {\it Phys. Rev.} B \textbf{84} 014436

\bibitem{hollm08} Hollaman N, Haverkort M W, Cwik M, Benomer M, Reuther M,
                   Tanaka A and Lorentz T 2008
                   {\it New J. Phys.} \textbf{10} 023018

\bibitem{Ole84} Ole\'s A M 1983
                   {\it Phys. Rev.} B \textbf{28} 327

\bibitem{Horsch} Horsch P 2007
                    {\it Orbital Physics in Transition-metal Oxides:
                    Magnetism and Optics}, in Handbook of Magnetism and
                    Advanced Magnetic Materials,
                    edited by Kronm\"uller H and Parkin S 2007,
                    Volume 1: {\it Fundamentals and Theory}, J. Wiley and Sons, Ltd.

\bibitem{Griffith} Griffith J S 1971
                   {\it The Theory on Transition Metal Ions},
                   Cambridge University Press

\bibitem{buene05} B\"unemann J, Gebhard F, Ohn T, Weiser S and Weber W 2005
                   {\it Gutzwiller-Correlated Wave Functions: Application to
                   ferromagnetic nickel}, in the Chapter: Parameters for the
                   Coulomb Interaction, arXiv.org, Cornell Univ. Library,
                   arXiv:cond-mat/0503332.

\bibitem{Ole12} Ole\'s A M 2012
                   J. Phys.: Condens. Matter \textbf{24} 313201

\bibitem{Hor11} Horsch P and Ole\'s A M 2011
                   {\it Phys. Rev.} B \textbf{84} 064429 \\
                Avella A, Horsch P and Ole\'s A M 2013
                   {\it Phys. Rev.} B \textbf{87} 045132 
                   
\bibitem{mizok95} Mizokawa T and Fujimori A 1995
                   {\it Phys. Rev.} B \textbf{51} 12880 \\
                Mizokawa T and Fujimori A 1996
                   {\it Phys. Rev.} B \textbf{54} 5368

\bibitem{wakis08} Wakisaka Y, Hirata S, Mizokawa T,
                   Suzuki Y, Miyazaki Y and Kajitani T 2008
                   {\it Phys. Rev.} B \textbf{78} 235107

\bibitem{zhang04} Zhang W, X Huang Q, Zhang W and Hu A 2004
                   {\it J. Appl. Phys.} \textbf{95} 6822

\bibitem{wu05}  Wu W B, Huang D J, Okamoto J, Tanaka A, Lin H J, Chou F C,
                   Fujimori A and Chen C T 2005
                   {\it Phys. Rev. Lett.} \textbf{94} 146402

\bibitem{lda1}  Lee K-W, Kune\v{s} J and Pickett W E 2004
                   {\it Phys. Rev.} B \textbf{70} 045104

\bibitem{lda2}  Tanaka A and Hu X 2003
                   {\it Phys. Rev. Lett.} \textbf{91} 257006

\bibitem{land06} Landron S and Lepetit M 2006
                   {\it Phys. Rev.} B \textbf{74} 184507

\bibitem{Sto80} Stollhoff G and Fulde P 1980
                 {\it J. Chem. Phys.} \textbf{73} 4548\\
                Stollhoff G 1996 {\it J. Chem. Phys.} \textbf{105} 227\\
                Fulde P 1991
                   {\it Electron Correlations in Molecules and Solids},
                   Springer Series in Solid State Sciences, Vol.
                   100, Springer Verlag, Berlin.

\bibitem{Ole87} Ole\'s A M, Zaanen J and Fulde P 1987
                   {Physica} B\&C \textbf{148} 260 \\
                Ole\'s A M and Grzelka W 1991
                   {\it Phys. Rev.} B \textbf{70} 9531

\bibitem{ni}    Ro\'sciszewski K and Ole\'s A M 2011
                   {\it J. Phys.: Condens. Matter} \textbf{23} 265601

\bibitem{Ole86} Ole\'s A M, Pfirsch F, Fulde P and B\"ohm M C 1986
                   {\it J. Chem. Phys.} \textbf{85} 5183 \\
                Ole\'s A M, Pfirsch F, Fulde P and B\"ohm M C 1987
                   {\it Z. Phys.} B \textbf{66} 359

\bibitem{Seo12} Seo H, Posadas A and Demkov A A 2012
                   {\it Phys. Rev.} B \textbf{86} 014430

\bibitem{Sbo09} Sboychakov A O, Kugel K I, Rakhmanov A L and Khomskii D I 2009
                   {\it Phys. Rev.} B \textbf{80} 024423

\bibitem{zaa88} Zaanen J, Paxton A T, Sepsen O and Andersen O K 1988
                   {\it Phys. Rev.} B \textbf{60} 2685.

\bibitem{Ole11} de Gennes P G 1960
                   {\it Phys. Rev.} B \textbf{118} 141 \\
                van den Brink J and Khomskii D I 1999
                   {\it Phys. Rev. Lett.} \textbf{82} 1016 \\
                Ole\'s A M and Feiner L F 2002
                   {\it Phys. Rev.} B \textbf{65} 052414 \\
                Daghofer M, Ole\'s A M and von der Linden W 2004
                   {\it Phys. Rev.} B \textbf{70} 184430 \\
                Ole\'s A M and Khaliullin G 2011
                   {\it Phys. Rev.} B \textbf{84} 214414

\bibitem{her08} Hertz J T, Huang Q, McQueen T, Klimczuk T, Bos J W G,
                   Viciu L and Cava R J 2008
                   {\it Phys. Rev} B \textbf{77} 075119

\bibitem{gas06} Gasparovic G, Ott R A, Cho J H, Chou F C, Chu Y,
                   Lynn J W and Lee Y S 2006
                   {\it Phys. Rev. Lett.} \textbf{96} 046403

\bibitem{ohi10} Ohira-Kawamura S, Nagata T, Takeda K and Yoshizawa H 2010
                   {\it Physica} C \textbf{470} S691

\bibitem{garba08} Garbarino G, Monteverde M, N\'u\~{n}ez-Regueiro M,
                   Acha C, Foo M L and Cava R J 2008
                   {\it Phys. Rev.} B \textbf{77} 064105

\bibitem{cwik09} Cwik M, Benomar M, Finger T, Sidis Y,
                   Senff D, Reuther M, Lorenz T and Braden M 2009
                   {\it Phys. Rev. Lett.} \textbf{102} 057201

\bibitem{yokoi05} Yokoi M, Moyoshi T, Kobayashi Y, Soda M, Yasui Y,
                   Sato M and Kakurai L 2005
                   {\it J. Phys. Soc. Jpn.} \textbf{74} 3046

\bibitem{hel06} Helme L M, Boothroyd A T, Coldea R, Prabakaran D,
                   Stunault A, McIntyre G J and Kernavanois N 2006
                   {\it Phys. Rev.} B \textbf{73} 054405

\bibitem{yok05} Yokoi M, Moyoshi T, Kobayashi Y, Soda M, Yasui Y,
                   Sato M and Kakurai K 2005
                   {\it J. Phys. Soc. Jpn.} \textbf{74} 3046

\bibitem{khal}  Khaliullin G, Koshibae W and Maekawa S 2004
                   {\it Phys. Rev. Lett.} \textbf{93} 176401

\bibitem{harr05} Harrison W A 2005
                   {\it Elementary Electronic Structure}, World Scientific, London



\end{thebibliography}
\end{document}